\documentstyle[12pt]{article}

\begin{document}

\thispagestyle{empty}
\setcounter{page}{0}

\vspace*{-8ex}
 \null \hfill MPI-PhT/96-26 \\
 \null \hfill hep-th/9605161 \\
 \null \hfill Mai 1996 \\[10mm]

\begin{center}

{\LARGE\bf
  A q\hspace{0.5mm}-Deformation of the Harmonic Oscillator\\[8ex]}

{\large A. Lorek\footnote{Supported by the
German-Israeli Foundation (G.I.F.)}$^{\,1}$, A. Ruffing$^{\,1}$ 
and J. Wess$^{1,2}$}
   \\ [7ex]

 $^1$ Max-Planck-Institut f\"{u}r Physik \\
   (Werner-Heisenberg-Institut)\\
 F\"{o}hringer Ring 6 , D - 80805 M\"{u}nchen, Germany \\
 Tel. (89) 32354-0, Fax (89) 3226704\\
 email: lorek@mppmu.mpg.de\\ [3ex]

 $^2$ Sektion Physik, Universit\"{a}t M\"{u}nchen \\
   Theresienstr.\,37, D - 80333 M\"{u}nchen, Germany \\ [16ex]

\end{center}

\begin{abstract}

\noindent
The q-deformed harmonic oscillator is studied in the light of q-deformed
phase space variables. This allows a formulation of the corresponding
Hamiltonian in terms of the ordinary canonical variables $x$ and $p$.\\
The spectrum shows unexpected features such as degeneracy and an additional 
part that cannot be reached from the ground state by creation operators.
The eigenfunctions show lattice structure, as expected.

\end{abstract}

\clearpage

\renewcommand{\theequation}{\arabic{section}.\arabic{equation}}
\newcommand{\nn}{\nonumber}
\newcommand{\be}{\begin{equation}}
\newcommand{\ee}{\end{equation}}
\newcommand{\bea}{\begin{eqnarray}}
\newcommand{\eea}{\end{eqnarray}}
\newcommand{\alb}{\bar{\alpha}}
\newcommand{\beb}{\bar{\beta}}
\newcommand{\ZZ}{\mbox{Z\hspace{-1.35mm}Z\hspace{1.35mm}}}
\newcommand{\NN}{\mbox{I\hspace{-0.58mm}N\hspace{0.58mm}}}
\newcommand{\CC}{\mbox{C\hspace{-2.1mm}l\hspace{2.1mm}}}
\newcommand{\RR}{\mbox{{\rm R}\hspace{-2.5mm}I\hspace{1mm}}}

\section*{Introduction}

In a previous paper \cite{pha} we have introduced the q-deformed phase
space and we have seen that it can be embedded in ordinary phase space.
Thus the dynamics of a Hamiltonian expressed in terms of q-deformed phase space
variables can be interpreted as the dynamics of a quantum mechanical 
Hamiltonian with complicated interactions but in the standard Hilbert space of 
quantum mechanics.\\
A. Macfarlane and L. Biedenharn \cite{mcf,bie} have studied the q-deformed
harmonic oscillator based on an algebra of q-deformed creation and 
annihilation operators. They have found the spectrum and eigenvalues of such
a harmonic oscillator under the assumption that there is a state with a
lowest energy eigenvalue.\\
In this paper we are going to express the
q-deformed creation and annihilation operators in terms of q-deformed phase
space variables and we obtain this way an ordinary Hamiltonian that can be
diagonalized due to our knowledge of the q-deformed system. The spectrum,
however, has a richer and more complicated structure than one would have
expected.
There is the part with a lowest eigenvalue belonging to the groundstate of
the system:
\[
E_m \,=\, \frac{1-q^{-2m}}{1-q^{-2}}\hspace{12mm}m=0,1,\cdots,\infty
\]
These eigenvalues are  bounded from above by
\[
E_{[\infty]} \,=\,\frac{\omega}{1-q^{-2}}
\]
Above $E_{[\infty]}$ there is an unbounded spectrum which has no lowest
eigenvalue:
\[
E_n \,=\, \omega \,(\frac{1}{1-q^{-2}} + \beta\beb\,q^{2n}) \hspace{10mm}
n=-\infty,\cdots,\infty
\]
 $\beta$ is related to the mass and the frequency of the undeformed oscillator.
The spectrum is twofold degenerate.
In the limit $q\rightarrow 1$ the unbounded part of the spectrum disappears.
The degeneracy of the bounded part is lifted as the support of the 
eigenfunctions is shifted away for half of the eigenstates, the other half
tends to the eigenfunctions of the ordinary oscillator. This will be explained
in detail in this paper.\\
The model shows a general feature of q-deformed dynamical systems. In the
neighbourhood ($q\ne 1$) of ordinary quantum mechanical systems we can expect 
a class of Hamiltonians that can be diagonalized if the ordinary Hamiltonian 
can be diagonalized. The parameter $h$ of $q=e^h$ can be interpreted as an 
interaction constant, however, the quantum mechanical problem is solved 
non-perturbatively. The interaction produces a richer spectrum and the 
eigenfunctions exhibit lattice-like structures.

In chapter 1 we define the model. In chapter 2 and 3 we diagonalize the 
Hamiltonian and find its eigenfunctions for the bounded and unbounded part 
of the spectrum. In chapter 4 we relate the model to a quantum 
mechanical model and show that it is a q-deformation of the undeformed 
harmonic oscillator. In chapter 5 we study the transition of the wave 
functions to those of the ordinary harmonic oscillator in the limit 
$q\rightarrow 1$. In chapter 6 we show that q-deformed Hermite polynomials 
arise in a natural 
way.

\section{The model}

\setcounter{equation}{0}

The q-deformed harmonic oscillator has first been studied by 
A. Macfarlane \cite{mcf} and L. Biedenharn \cite{bie}. They introduced the 
algebra of q-deformed creation and annihilation operators and they computed the
spectrum of the corresponding Hamiltonian, $H=\omega a^+ a$, assuming the
existence of a ground state.\\
We shall express the creation and annihilation operators in terms of 
q-deformed phase space variables \cite{pha}:
\be
\begin{array}{c}
q^{\frac{1}{2}}\,X\,P - q^{-\frac{1}{2}}\,P\,X = i\,U \\[4mm]
U\,X = q^{-1}\,X\,U \;\;\; , \;\;\;\;\; U\,P = q\,P\,U
\end{array}
\label{e1}
\ee
We assume $q$ to be real and $q>1$.\\
The algebra (\ref{e1}) allows $X$ and $P$ to be hermitean and $U$ unitary:
\be
X^+ = X \;\; , \;\;\;P^+ = P \;\; , \;\;\; U^+ = U^{-1}
\label{e2}
\ee
The operators
\bea
a   &=& \alpha\,U^{-2M} + \beta\,U^{-M}\,P \nn \\
a^+ &=& \:\alb\;U^{2M}\: + \;\beb\,P\,U^{M}  
\label{e3}
\eea
with $M \in \NN$, $\alpha,\beta \in \CC$, satisfy the algebra
\be
a a^+ - q^{-2M} a^+ a = (1-q^{-2M})\; \alpha \alb = 1
\label{e4}
\ee
The right hand side can be normalized to 1 for $M>0$, $q>1$. This determines 
$\alpha$ up to a phase. In chapter 4 we shall relate $\beta$ to the 
mass and the frequency of the undeformed harmonic oscillator and we shall 
see that $\alpha \beb = \alb \beta$.\\
The operators $a$ and $a^+$ play the role of annihilation and creation
operators (for $M<0\, ,\:$ $a$ and $a^+$ would change role and we would deal
with the same algebra). A more general expression for $a$ and $a^+$ in terms
of $P$, $X$ and $U$ has been studied in ref. \cite{ruf} and by J. Seifert
\cite{sei}.\\
From (\ref{e4}) follows that $a$ lowers an energy eigenvalue $E$ of
the Hamiltonian \linebreak 
$H=\omega a^+ a$ as long as 
\[
E < \frac{\omega}{1-q^{-2M}} \,\equiv\,E_{[\infty]}  \;\;\;\;\; .
\]

For $E > E_{[\infty]}$, $a$ and $a^+$ change their role, $a$ raises and
$a^+$ lowers the energy. Starting from an arbitrary eigenvalue 
$E_0 >  E_{[\infty]}$, we find a spectrum
\be
E_{\rm m} = \omega \left( q^{2{\rm m}M} E_0 
            + \frac{1-q^{2{\rm m}M}}{1-q^{-2M}} \right)
\label{e9}
\ee
\vspace{-4mm}
\[
{\rm m} = -\infty,\, \ldots,\,  +\infty
\]
and
\bea
|E_{m-1}\rangle &=& \sqrt{\frac{\omega}{E_{m-1}}}\;a^+\,|E_m\rangle\nn\\
|E_{m+1}\rangle &=& \sqrt{\frac{\omega}{E_m}}\;a\,|E_m\rangle
\eea
For ${\rm m} \rightarrow -\infty$, the eigenvalue approaches $ E_{[\infty]}$,
independent of $E_0$. This leads to a representation that does not have a 
"lowest weight" state. Such representations have been studied in ref. 
\cite{dyn} as well.\\
For $E=E_{[\infty]}$, $a$ and $a^+$ do not change the eigenvalue.

That we have to encounter eigenvalues  $E > E_{[\infty]}$ follows from the 
following short argument.\\
The representations of the algebra (\ref{e1}) have been studied in ref. 
\cite{pha} and we know that $P$ is an unbounded operator:
\be
P |n,\sigma\rangle = \sigma q^n \;|n,\sigma\rangle
\label{ee5}
\ee
\vspace{-4mm}
\[
\sigma = +,- \;\;\; , \;\;\;\;\;   n = -\infty,\, \ldots,\,  +\infty
\]
From (\ref{e3}) follows that $a^+ a$ cannot be bounded either:
\be
\langle n,\sigma| a^+ a |n,\sigma\rangle = \alpha\alb + \beta\beb\,q^{2n}
\;\;\;\;\;\; .
\label{e11}
\ee

\section{The bounded spectrum}

\setcounter{equation}{0}

By an explicit construction we show that the Hamiltonian $H=\omega a^+ a$ 
has eigenvalues 
$E < E_{[\infty]}$. In this energy range the operator $a$ lowers 
the energy eigenvalue. As the eigenvalues of $H$ cannot become negative 
we have to assume the existence of a "ground" state:
\be
a\, |0\rangle^{(M)} \,=\, 0 \hspace{2cm} 
\mbox{}^{(M)}\langle0\,|\,0\rangle^{(M)} \,=\, 1
\label{e5}
\ee
The spectrum follows immediately:
\be
a^+ a\; |{\rm n}\rangle^{(M)} \,=\, \frac{1-q^{-2{\rm n}M}}{1-q^{-2M}}\; 
|{\rm n}\rangle^{(M)} 
\,=\, [{\rm n}]_M \,|{\rm n}\rangle^{(M)}
\label{e6}
\ee
\vspace{-4mm}
\[
{\rm n} = 0,\, \ldots,\,  +\infty
\]
where we have defined the q-number $[{\rm n}]_M$. The normalized states are:
\be
|{\rm n}\rangle^{(M)} = \frac{1}{\sqrt{[{\rm n}]_M!}}\; 
\left( a^+ \right)^{\rm n} |0\rangle^{(M)} 
\label{e7}
\ee
\vspace{-2mm}
\[
\mbox{}^{(M)}\langle{\rm n}|{\rm m}\rangle^{(M)} = \delta_{{\rm n,m}}
\]
This spectrum is bounded from above:
\be
[{\rm n}]_M \;\stackrel{{\rm n}\rightarrow\infty}{\longrightarrow}\;
\frac{1}{1-q^{-2M}} =: [\infty]_M
\label{e8}
\ee
and explains our notation $E_{[\infty]}$.

We now construct a ground state by first expanding it in terms of the 
"momentum" eigenstates of (\ref{ee5}): 
\be
|0\rangle^{(M)}_{(\sigma)} = \sum^{+\infty}_{m=-\infty} \;c_m^{(\sigma)}\, 
|m,\sigma\rangle
\label{e12}
\ee
From (\ref{e5}) follows the recursion formula
\be
c_m^{(\sigma)} = -\sigma \;\frac{\alpha}{\beta}\; q^{-m}\,c^{(\sigma)}_{m-M}
\label{e13}
\ee
with $2M$ independent solutions. We define
\be
\tilde{c}_m^{(\sigma,\mu)} = c^{(\sigma)}_{mM+\mu} \hspace{20mm} 0\le \mu <M
\;, \;\;\mu \in \NN_0
\label{e14}
\ee
and find from (\ref{e13})
\be
\tilde{c}_m^{(\sigma,\mu)} = (-\sigma \frac{\alpha}{\beta})^m\,
q^{-\frac{1}{2}[Mm^2+Mm+2\mu m]}\,c_0^{(\mu)}
\label{e15}
\ee
The $2M$ independent ground states are
\be
|0\rangle_{(\sigma,\mu)}^{(M)} = \sum^{+\infty}_{m=-\infty} 
\tilde{c}_m^{(\sigma,\mu)}\; |mM+\mu,\sigma\rangle
\label{e16}
\ee
\vspace{-4mm}
\[
0\le \mu <M \;, \;\;\;\; \sigma=+,- \;\;\; .
\]
Due to the factor $q^{-\frac{1}{2}Mm^2}$ they are all of finite norm.
On each of these ground states we can act with the creation operator $a^+$ to
obtain exited states. The spectrum will be $2M$ times degenerate.

It is easy to compute the ground state expectation value of the momentum
operator:
\be
\mbox{}^{(M)}_{(\sigma,\mu)}\langle 0|P|0\rangle^{(M)}_{(\sigma,\mu)} 
\,=\, \sigma \sum^{+\infty}_{m=-\infty} \left( \frac{\alpha\alb}{\beta\beb}
\right)^m
q^{-Mm^2-2\mu m+\mu} \,|c_0^{(\mu)}|^2
\label{e17}
\ee
It is different from zero and the sign depends on $\sigma$.\\
The exited states can be obtained by acting with $a^+$ on the ground states.
For the first exited state we obtain:
\be
\mbox{}^{(M)}_{(\sigma,\mu)}\langle 1|P|1\rangle^{(M)}_{(\sigma,\mu)}
\,=\, \frac{2}{q^M(1+q^M)} \,\; \;
\mbox{}^{(M)}_{(\sigma,\mu)}\langle 0|P|0\rangle^{(M)}_{(\sigma,\mu)} 
\label{e18}
\ee
The expectation value of the momentum operator decreases in absolute value.

In analogy to ordinary quantum mechanics the eigenstates can be 
found by solving the recursion formula associated with the eigenvalue problem:
\be
H\,|E\rangle \,=\, E \,|E\rangle 
\label{e19}
\ee
\vspace{-3mm}
\[
H \,=\, a^+ a \hspace{4mm},\hspace{6mm} 
|E\rangle \,=\, \sum^{+\infty}_{n=-\infty} c_n \,
|n,\sigma\rangle \]
For simplicity we treat the case $M=1$, $\omega=1$. The respective 
recursion formula is:
\be
E\,c_n = (\alpha\alb + \beta\beb q^{2n})\,c_n 
   + \alpha\beb\,\sigma\,(q^n c_{n-1} + q^{n+1} c_{n+1}) 
\label{e20}
\ee
where we used $\alb\beta = \alpha\beb$.\\
The solution (\ref{e16}) with (\ref{e15}) for $M=1$ suggests the following 
ansatz:
\be
c_n = f_n \,(-\frac{\alpha}{\beta})^n\,q^{-\frac{1}{2}(n^2+n)}
\label{e23}
\ee
The recursion formula for the $f_n$'s is:
\be
f_{n+1} = -\frac{1}{\alpha\alb}\,\left( E-\alpha\alb -\beta\beb q^{2n}\right) 
\,f_n - \frac{\beta\beb}{\alpha\alb}\,q^{2n}\,f_{n-1}
\label{e24}
\ee
Following our knowledge of quantum mechanics we try a polynomial ansatz:
\be
f_n^{(K)} = \sum_{r=0}^{K} \, \eta_r\,q^{-2nr}
\label{e25}
\ee
From (\ref{e24}) follows:
\be
\frac{\beta\beb}{\alpha\alb} \left(1-q^{2(l+1)}\right)\,\eta_{l+1} = 
\left( q^{-2l} + \frac{E-\alpha\alb}{\alpha\alb}\right)\,\eta_l
\label{e26}
\ee
\vspace{-4mm}
\[ l = -1,\,\ldots,\, K \]
For $l=K$ we learn
\be
E \,=\, (1-q^{-2K})\,\alpha\alb\, =\,  \frac{1-q^{-2K}}{1-q^{-2}}
\label{e27}
\ee
These are our energy eigenvalues for $E<E_{[\infty]}=\alpha\alb$. 
They are determined from
the requirement that the series (\ref{e25}) terminates. The other values of
$l$ determine $\eta_{l+1}$ in terms of $\eta_l$. We find $\eta_{-1}=0$ in
agreement with the ansatz (\ref{e25}).\\
If we did not make a polynomial ansatz, eqn. (\ref{e26}) would tell us
the asymptotic behaviour of the series (\ref{e25}):
\be
\frac{\eta_{l+1}}{\eta_l}\; \stackrel{l\rightarrow\infty}{\longrightarrow}\;
\frac{\alpha\alb - E}{q^{2l}\beta\beb q^2}
\label{e28}
\ee
This is the behaviour of a series of the form $\eta_l \sim q^{-l^2}$. Such
a series can be summed and behaves like $q^{n^2}$. This would lead to
non-normalizable states. So our ansatz (\ref{e23}) with (\ref{e25}) does not 
lead to the eigenstates for $E \ge E_{[\infty]}$.

For $E = E_{[\infty]} = \alpha\alb $, the recursion formula (\ref{e20}) is 
identical to
the one of ref. \cite{heb} (eqn. (15)) for coefficients $d_n = (-iq)^n c_n$. 
There it was shown that the recursion formula leads to a unique solution with 
the corresponding state normalizable. This solution can be expressed in terms 
of q-deformed cosine and sine functions.\\
For $E \ne \alpha\alb$ the recursion formula (\ref{e20}) has terms 
tending to infinity for $n\rightarrow +\infty$ and $n\rightarrow -\infty$.
For $n\rightarrow +\infty$, (\ref{e20}) tends asymptotically to the recursion
formula of ref.\cite{heb} with the above substitution $d_n = (-iq)^n c_n$. 
The same is true for $n\rightarrow-\infty$ with the identification 
$d_n= i^n c_{-n}$. This determines the asymptotic behaviour of the exact 
solution of (\ref{e20}) for $n\rightarrow \pm \infty$ which will only match 
for certain values of $E$.\\ 
That these solutions of the recursion formula exist will be shown in the 
following chapter.

\section{The unbounded spectrum}

\setcounter{equation}{0}

To compute the spectrum above $E_{[\infty]}$ we retreat to 
perturbation theory. The simple form of the interaction will allow us to
draw exact conclusions. In the following we set $M=1$ and $\omega=1$.
We will only treat the case $\sigma=+1$ (\ref{ee5}), the case $\sigma=-1$
is analogous.\\
The complete Hamiltonian is 
\be
H = \alpha\alb + \beta\beb\,P^2 + \alpha\beb \left(UP+PU^{-1}\right)
\label{d1}
\ee
of which we will treat the second part as a perturbation:
\bea
H_I &=& \alpha\beb \left(UP+PU^{-1}\right) \label{d2}\\
H_0 &=& \alpha\alb + \beta\beb\,P^2  \nn
\eea
$H_0$ is diagonal in the momentum representation:
\be
H_0\,|m\rangle = \left(\alpha\alb + \beta\beb\,q^{2m}\right)\,|m\rangle
\equiv E_m^{(0)}\,|m\rangle
\label{d3}
\ee
As $\alpha\alb = \frac{1}{1-q^{-2}} = E_{[\infty]}$ we see that all the 
unperturbed energy eigenvalues are larger than $E_{[\infty]}$.

It is obvious from (\ref{d2}) that the first order correction $E_m^{(1)}$
to the energy eigenvalue is zero. An explicit calculation shows that this
is also true for the second order correction $E_m^{(2)}$. We shall prove by
induction that the corrections to the energy eigenvalues are zero to all 
orders. \\
The standard expression in perturbation theory \cite{schiff} for $E_m^{(l)}$, 
with the assumption that $E_m^{(r)} = 0$ for $r<l$, becomes:
\be
E_m^{(l)} = \sum_{\stackrel{\scriptstyle n_1,n_2,\dots,n_{l-1}}{n_r\ne m}} \;
\frac{\langle m|H_I|n_{l-1}\rangle\langle n_{l-1}|H_I|n_{l-2}\rangle \cdots
\langle n_1|H_I|m\rangle }{(E_m-E_{n_{l-1}})(E_m-E_{n_{l-2}}) \cdots 
(E_m-E_{n_1})}
\label{d4}
\ee
This sum is restricted to $n_r\ne m$.\\
The nonvanishing matrix elements of $H_I$ are
\bea
\langle n+1|H_I|n\rangle &=& \alpha\beb\,q^{n+1} \label{d5}\\
\langle n-1|H_I|n\rangle &=& \alpha\beb\,q^n \nn
\eea
As this restricts the jumps in $n$ to $\Delta n=\pm 1$ we conclude from the
restriction  $n_r\ne m$ that the sum (\ref{d4}) splits in a natural way into
two parts  $n_r>m$ for all $r$ and  $n_r<m$. We shall show that these two 
contributions are equal in magnitude and opposite in sign.\\
For any "path"  $n_r=m+k_r$ ($r=1,\dots,l-1$) with all $k_r>0$ there is a path
$n'_r=m-k_r$. We compare their contributions to (\ref{d4}). First we notice
that, due to (\ref{d5}), $l$ has to be even. This leads to an odd number of
factors in the denominator and an even number of factors in the numerator.
From the contributions to the path we separate the first and the last step in
the two cases:
\be
\frac{\langle m|H_I|m+1\rangle\langle m+1|H_I|m\rangle}{E_m-E_{m+1}}
= \alpha\alb \frac{q^2}{1-q^2}
\label{d6}
\ee
and
\be
\frac{\langle m|H_I|m-1\rangle\langle m-1|H_I|m\rangle}{E_m-E_{m-1}}
= -\alpha\alb \frac{q^2}{1-q^2}
\label{d7}
\ee
The rest of the factors will give the same contribution in both cases:
\bea
&&\frac{\langle n_{l-1}|H_I|n_{l-2}\rangle \cdots 
\langle n_2|H_I|n_1\rangle}{(E_m-E_{n_{l-2}}) \cdots (E_m-E_{n_1})} \nn\\
&=&\left(\frac{\alpha}{\beta}\right)^{l-2} \frac{q^{n_1+n_2+\cdots+n_{l-2}+
\frac{1}{2}(l-2)}}{q^{2m(l-2)}(1-q^{2(n_1-m)}) \cdots (1-q^{2(n_{l-2}-m)})}
\label{d8}\\
&=&\left(\frac{\alpha}{\beta}\right)^{l-2}q^{-m(l-2)} 
\frac{q^{k_1+k_2+\cdots+k_{l-2}+\frac{1}{2}(l-2)}}{(1-q^{2k_1})
 \cdots (1-q^{2k_{l-2}})}\nn
\eea
The factor $q^{n_1+n_2+\cdots+n_{l-2}+\frac{1}{2}(l-2)}$ results from the
following consideration. Looking at (\ref{d5}) we see that starting from
a state $|n_j\rangle$ at the right hand side we get factors 
$q^{n_j+1}$ or $q^{n_j}$ depending whether the state on the left hand side
is $|n_j+1\rangle$ or $|n_j-1\rangle$. In our path we have to have as many
"increasing" as "decreasing" matrix elements. For each of the 
$\frac{1}{2}(l-2)$ "increasing" steps (the first and last step we treated 
separately) we have a factor of $q$ in addition to the factor $q^{n_j}$.\\
Now we look at the corresponding contribution from the path $n'_r = m-k_r$:
\bea
&&\frac{\langle n'_{l-1}|H_I|n'_{l-2}\rangle \cdots 
\langle n'_2|H_I|n'_1\rangle}{(E_m-E_{n'_{l-2}}) \cdots (E_m-E_{n'_1})} \nn\\
&=&\left(\frac{\alpha}{\beta}\right)^{l-2} \frac{q^{n'_1+n'_2+\cdots+n'_{l-2}+
\frac{1}{2}(l-2)}}{q^{2m(l-2)}(1-q^{2(n'_1-m)}) \cdots (1-q^{2(n'_{l-2}-m)})}
\label{d9}\\
&=&\left(\frac{\alpha}{\beta}\right)^{l-2}q^{-m(l-2)} 
\frac{q^{-k_1-k_2-\cdots-k_{l-2}+\frac{1}{2}(l-2)}}{(1-q^{-2k_1})
 \cdots (1-q^{-2k_{l-2}})}\nn
\eea
(\ref{d8}) and (\ref{d9}) give the same contribution. There
is no change in sign as in (\ref{d6}) and (\ref{d7}) because there is an 
even number of factors in the denominator. This completes our proof by 
induction and we get the surprising result
\be
E_m = E_m^{(0)} = \alpha\alb + \beta\beb \,q^{2m}
\label{d10}
\ee
It determines $E_0$ in (\ref{e9}):
\be
E_0 = \alpha\alb + \beta\beb 
\label{d11}
\ee

After having found the exact energy eigenvalues we will now proceed to
calculate the corresponding eigenstates $|E_m\rangle$
\be
|E_m\rangle \,= \sum_{n=-\infty}^{+\infty} \,c_n^{(m)}\,|n\rangle
\label{t1}
\ee
where $|n\rangle$ are the momentum eigenstates and the eigenstates of the
unperturbed Hamiltonian as well. As usual in perturbation theory, the
coefficients $c^{(m)}_n$ are expanded
\be
c_{m+k}^{(m)} \,=\, \sum_l \,a_{l,k}^{(m)} \hspace{10mm} \mbox{\rm for}\;\;k\ne 0
\;\;,\hspace{10mm}c^{(m)}_m = 1
\label{t2}
\ee
After having shown that the energy corrections are zero we have the following
expression for the expansion:
\be
a_{l,k}^{(m)}\, =\! \sum_{\stackrel{\scriptstyle n_1,n_2,\dots,n_{l-1}}{n_r\ne m}}
\;\frac{\langle m+k|H_I|n_{l-1}\rangle\langle n_{l-1}|H_I|n_{l-2}\rangle \,
\cdots\, \langle n_1|H_I|m\rangle }{(E_m-E_{m+k})(E_m-E_{n_{l-1}})\,\cdots\, 
(E_m-E_{n_1})}
\label{t3}
\ee
For $k>0$ we have $n_r>m$ for all $r$, for $k<0$ we have $n_r<m$. Each term
of the sum is composed of $l$ factors of the type
\be
\frac{\langle n|H_I|n'\rangle}{E_m-E_n}\, =\, \frac{\alpha\beb \,
\left(q^n\,\delta_{n,n'+1} + q^{n+1}\,\delta_{n,n'-1}\right)}{\beta\beb\,q^{2m}
(1-q^{2(n-m)})}
\label{t4}
\ee
We first take care of the case $k>0$. There we have the following estimate for
the factors:
\be
\left| \,\frac{\langle n|H_I|n'\rangle}{E_m-E_n}\,\right| 
\;\,\le\,\; |\frac{\alpha}{\beta}|\: q^{-m}\; \frac{1}{1-q^{-2}}
\label{t5}
\ee
There are $l$ factors of this type in each term of the sum (\ref{t3}). There
certainly cannot be more than $2^l$ terms in the sum. Remember that at each of
the $l$ steps in the "path" we can at most go one step up or one step down.\\
This now leads to the estimate
\be
\left| \,a_{l,k}^{(m)} \,\right| \;\le\;  |\frac{\alpha}{\beta}|^l q^{-ml}\: 
\frac{1}{(1-q^{-2})^l}\; 2^l
\label{t6}
\ee
Furthermore we see from (\ref{t3}) that
\be
a_{l,k}^{(m)} = 0 \;\;\;\;\;\; \mbox{\rm for} \;\;\;\;l<k
\label{t7}
\ee
Now we get an upper bound for $c_{m+k}^{(m)}$ for $k>0$:
\be
\left|\, c_{m+k}^{(m)}\, \right| \;\,\le\,\; (\, 2 \,|\frac{\alpha}{\beta}|\, 
\frac{q^{-m}}{1-q^{-2}})^k
\;\;\sum_{r=0}^{\infty}\, ( \,2\,|\frac{\alpha}{\beta}|\,
\frac{q^{-m}}{1-q^{-2}})^r
\label{t8}
\ee
For fixed value of $\alpha$ and $\beta$ we can always find an $m$ such that
\be
 2\:|\frac{\alpha}{\beta}|\; \frac{q^{-m}}{1-q^{-2}} \:\;<\:\; 1
\label{t9}
\ee
and the sum in (\ref{t8}) converges.

We could make a similar estimate for the case $k<0$. By a general analysis
of the recursion formula (\ref{e20}) we will rather relate this case to the
case $k>0$.\\
We substitute (\ref{d10}) for the energy eigenvalues in (\ref{e20}) and obtain
a recursion formula for the coefficients $c_n^{(m)}$:
\be
\beta\beb\,\left(q^{2m}-q^{2n}\right)\,c_n^{(m)}\: =\: \alpha\beb \,\left(
q^n\, c_{n-1}^{(m)} + q^{n+1}\, c_{n+1}^{(m)}\right)
\label{t10}
\ee
This can be seen as a recursion for decreasing or increasing indices:
\bea
c^{(m)}_{m+k} &=& \frac{\beta}{\alpha}\, q^{m-k} \left(1-q^{2(k-1)}\right)
c^{(m)}_{m+k-1} - q^{-1}\,c^{(m)}_{m+k-2} \label{t11}\\
c^{(m)}_{m-k} &=& \frac{\beta}{\alpha}\, q^{m+k-1} \left(1-q^{-2(k-1)}\right)
c^{(m)}_{m-(k-1)} - q \,c^{(m)}_{m-(k-2)} \nn
\eea
If we substitute $c^{(m)}_{m-k} = (-q)^k \,\tilde{c}^{(m)}_{m+k}$ in the
second relation we see that $\tilde{c}^{(m)}_{m+k}$ satisfies the first
relation.\\
Our perturbation expansion (\ref{t2}) is normalized by $c^{(m)}_m=1$.
From (\ref{t11}) follows
\be
c^{(m)}_{m+1} = -q^{-1} c^{(m)}_{m-1} 
\label{t12}
\ee
Thus $c^{(m)}_{m+k}$ and $\tilde{c}^{(m)}_{m+k}$ not only satisfy the same
recursion relation but also have the same initial values. Therefore we obtain:
\be
c^{(m)}_{m-k} = (-q)^k \,c^{(m)}_{m+k}
\label{t13}
\ee
The previous estimate (\ref{t8}) now tells us that
\be
|\, c_{m-k}^{(m)} \,| \;\le\; (\, 2\,q\, |\frac{\alpha}{\beta}|\, 
\frac{q^{-m}}{1-q^{-2}})^k
\;\sum_{r=0}^{\infty}\, (\, 2\,|\frac{\alpha}{\beta}|
\,\frac{q^{-m}}{1-q^{-2}})^r
\label{t14}
\ee
We see that the state $|E_m\rangle$ is normalizable for
\be
2\,q\, |\frac{\alpha}{\beta}|\, \frac{q^{-m}}{1-q^{-2}} \;<\; 1
\label{t15}
\ee
This of course is a very rough estimate but it proves the existencs of
normalizable states for $m$ large enough. The other energy eigenstates of
the unbounded spectrum can be reached by applying the operator $a^+$.

Perturbation theory can be used much more efficently to give the exact
coefficients $c^{(m)}_n$ in the expansion of the eigenstates (\ref{t1}). 
This will be shown now. We first exhibit the 
$m$-dependence of the coefficients $c^{(m)}_n$ by having a closer look at
(\ref{t3}) and (\ref{t4}). As in the study of the energy eigenvalues it is 
useful to label the path by $n_r=m+k_r$ as in (\ref{d8}). Equation (\ref{t4})
becomes
\be
\frac{\langle m+k_{r+1}|H_I|m+k_r\rangle}{E_m-E_{m+k_{r+1}}} \;=\;
(\frac{\alpha}{\beta}\, q^{-m})\,\frac{q^{k_{r+1}}\left(\delta_{k_{r+1},k_r+1} 
+ q\,\delta_{k_{r+1},k_r-1}\right)}{1-q^{2k_{r+1}}}
\label{t16}
\ee
The $m$-dependence is entirely contained in the factor 
$\frac{\alpha}{\beta}q^{-m}$. We obtain from (\ref{t3})
\be
a^{(m)}_{l,k}\; =\; (\frac{\alpha}{\beta}\,q^{-m})^l\,f_{l,k}(q)
\label{t17}
\ee
From (\ref{t2}) follows
\be
c_{m+k}^{(m)} = F_k(\frac{\alpha}{\beta}\,q^{-m}) \;\;\;\;, \;\;\;\;
c^{(m)}_m= F_0(\frac{\alpha}{\beta}\,q^{-m}) = 1
\label{t18}
\ee
or
\be
|E_m\rangle = \sum_{k=-\infty}^{\infty} F_k(\frac{\alpha}{\beta}\,q^{-m})
\;|m+k\rangle
\label{t19}
\ee
We now use the fact that the operator $a$ applied to this state leads to a
state with eigenvalue $E_{m+1}$. The normalization of our states is fixed
by the condition $F_0(\frac{\alpha}{\beta}q^{-m}) = c^{(m)}_m = 1$.
\bea
a\,|E_m\rangle &=& \sum_{n=-\infty}^{\infty} \left(\alpha\, c^{(m)}_{n-2} 
+\beta \,q^{n-1}\,c^{(m)}_{n-1} \right) \,|n\rangle\label{t20}\\
&=& \gamma_m\,|E_{m+1}\rangle
    \;=\; \gamma_m \, \sum_{n=-\infty}^{+\infty} \,c_n^{(m+1)}\,|n\rangle\nn
\eea
where $\gamma_m$ is a relative normalization factor.\\
From (\ref{t20}) follows
\be
\gamma_m c^{(m+1)}_n = \alpha \,c^{(m)}_{n-2} + \beta\, q^{n-1}\,c^{(m)}_{n-1}
\label{t21}
\ee
As we know that $c^{(m+1)}_{m+1}=1$ we find
\be
\gamma_m  = \alpha\, c^{(m)}_{m-1} + \beta\, q^m
\label{t22}
\ee
We are now going to combine the relations (\ref{t10}), (\ref{t21}) and 
(\ref{t22}). The recursion relation (\ref{t10}) can be written in the form
\be
\alpha\, c^{(m)}_{n-2} + \beta\, q^{n-1}\,c^{(m)}_{n-1} =
\beta \,q^{2m-(n-1)}\,c^{(m)}_{n-1} -\alpha\, q\, c^{(m)}_n
\label{t23}
\ee
From equation (\ref{t21}) we find
\be
\gamma_m\, c^{(m+1)}_n = \beta\, q^{2m-(n-1)}\,c^{(m)}_{n-1} -\alpha\, q\,
 c^{(m)}_n
\label{t24}
\ee
Setting $n=m$ and using the explicit $m$-dependence of the $c^{(m+l)}_n$ when
expressed throught the functions $F_k$ (\ref{t18}) we obtain an expression
for $F_{-1}$
\be
\left(\,\alpha \,F_{-1}(\frac{\alpha}{\beta}\,q^{-m}) + \beta\, q^m\,\right)\,
F_{-1}(\frac{\alpha}{\beta}\,q^{-(m+1)}) 
= \beta\, q^{m+1}\, F_{-1}(\frac{\alpha}{\beta}\,q^{-m}) - \alpha\,q
\label{t25}
\ee
This is an equation that can be solved for $F_{-1}$, at least perturbatively.
We abbreviate $z=\frac{\alpha}{\beta}q^{-m}$:
\be
z\,F_{-1}(z)\,F_{-1}(q^{-1}z)=q\left( F_{-1}(z) - q^{-1} F_{-1}(q^{-1}z)\right)
-qz
\label{t26}
\ee
If we make an ansatz
\be
F_{-1}(z) = \sum_{k=0}^{\infty} d_k\,z^k \;\;\;\;\;\; ,
\label{t27}
\ee
equation (\ref{t26}) allows us to compute the coefficients $d_k$. We find that
$d_{2k}=0$ and the first $d$'s for odd $k$ are
\bea
d_1 &=& \frac{1}{1-q^{-2}} \nn\\[-1mm]
d_3 &=& \frac{q^{-2}}{(1-q^{-2})^2(1-q^{-4})} \label{t28}\\
d_5 &=& \frac{q^{-3}(q^{-1}+q^{-3})}{(1-q^{-2})^3(1-q^{-4})(1-q^{-6})}\nn
\eea
This reproduces in a very compact way the result one obtains for
the coefficient $c^{(m)}_{m-1}$ from (\ref{t2}) and (\ref{t3}). The connection
is
\be
F_{-1}(\frac{\alpha}{\beta}\,q^{-m})\;=\; c^{(m)}_{m-1} \;=\; 
\sum_{l=0}^{+\infty}a_{-1,l}^{(m)}
\ee
\vspace{-2mm}
\[
a^{(m)}_{-1,k} \;=\; d_k\,(\frac{\alpha}{\beta}\,q^{-m})^k
\]
The other coefficients $c^{(m)}_n$ can now be calculated from the recursion
formula (\ref{t10}).

\section{The model and its relation to the undeformed harmonic oscillator}

\setcounter{equation}{0}

In this chapter we are going to show that the Hamitonian of the previous
chapters can be interpreted as a q-deformation of the usual harmonic 
oscillator.
We shall express the q-deformed phase space variables $X$, $P$ and $U$ in
terms of the usual canonical variables as we did in ref. \cite{pha}:
\be
P = \hat{p} \hspace{20mm} U = q^{-\frac{i}{2}(\hat{x}\hat{p}+\hat{p}\hat{x})}
\label{f1}
\ee
with
\vspace{-4mm}
\be
[\hat{x},\hat{p}] = i \;\; , \;\;\; \hat{x}^+ = \hat{x} \;\; , \;\;\;
 \hat{p}^+ = \hat{p} 
\label{f2}
\ee
As a consequence of (\ref{f2}), $P$ and $U$ will satisfy (\ref{e1}) and
(\ref{e2}).\\
We realize the relations (\ref{f2}) as follows:
\be
\hat{x} = x \;\;\; , \;\;\;\; \hat{p} = p + \frac{1}{\sqrt{1-q^{-2M}}}\;\gamma
\;\;\; , \;\;\;\;\;\; \gamma \in \RR
\label{f3}
\ee
\vspace{-4mm}
\[
p = -i\,\frac{\partial}{\partial x}\hspace{3cm}\mbox{}
\]
We now insert (\ref{f1}) into (\ref{e3}) and study the limit $q\rightarrow 1$,
which for $q=e^h$ means $h\rightarrow 0$.\\
The constant $\alpha$ of (\ref{e3}) is singular
\be
\alpha = \frac{e^{i\phi}}{\sqrt{1-q^{-2M}}} 
\approx \frac{e^{i\phi}}{\sqrt{2Mh}}
\label{f4}
\ee
This shows that $\sqrt{h}$ is a natural expansion parameter. In (\ref{f3}) 
a similar singular factor occurs for $\hat{p}$. The two singularities have
to conspire to produce a finite result. We expand:
\bea
a &=& \frac{e^{i\phi}}{\sqrt{2Mh}}\;\left( 1+ i\gamma x \sqrt{2Mh}\right)
\label{f5}\\
 & & +\, \beta \left(1+\frac{1}{2}i\gamma x \sqrt{2Mh}\right)
 \left( p + \frac{\gamma}{\sqrt{2Mh}} \right) + O(\sqrt{h})\nn
\eea
The singular terms cancel if:
\be
\beta\,\gamma = - e^{i\phi}
\label{f6}
\ee
Recall that $\gamma$ has to be real and therefore $\alpha\beb = \alb\beta$.\\
We find:
\be
a \;\stackrel{h\rightarrow 0}{\longrightarrow}\; e^{i\phi}
\left(\frac{1}{2}\,i\gamma x - \frac{1}{\gamma}\,p\right) = a_0
\label{f7}
\ee
There are two possibilities to identify (\ref{f7}) with the usual annihilation
operator of the harmonic oscillator:
\be
e^{i\phi} = \mp i \;\;\; ,\;\;\;\; \gamma = \pm \sqrt{2m\omega} \;\;\; ,
\;\;\;\; \beta = \frac{i}{\sqrt{2m\omega}}
\label{f8}
\ee
The relevant difference of these two choices is the sign of $\gamma$. The
constants $m$, $\omega$ are the mass and frequency of the unperturbed 
oscillator.

\noindent
The next term in the expansion is easily computed:
\be
a = a_0 \mp \frac{i}{4} \sqrt{2Mh} \;(1-\frac{3}{2}\gamma^2 x^2) + \cdots
\label{f9}
\ee
The two signs refer to the two choices in (\ref{f8}).\\
For the Hamiltonian we compute:
\be
H = a_0^+ a_0 \mp \frac{1}{4} \sqrt{\frac{Mh}{m\omega}} 
\left( p\,(1-3m\omega x^2) + (1-3m\omega x^2)\,p \right)
\label{f10}
\ee
The Hamiltonian $H=\omega a^+ a$ can be viewed as a quantized harmonic
oscillator with specific interactions where $\sqrt{h}$ appears as a coupling
constant. In this sense it represents a q-deformation of the harmonic
oscillator. It is interesting to write the full Hamiltonian in terms of
the variables $x$ and $p$:
\bea
 H &=& \omega \left( \frac{1}{1-q^{-2M}} + \frac{1}{2m\omega} 
\left(p \pm \sqrt{\frac{2m\omega}{1-q^{-2M}}}\right)^2 \;\right)\label{f11} \\
&& -\, \sqrt{\frac{\omega}{2m(1-q^{-2M})}}\;
\left[ q^{-\frac{i}{2}M(xp+px)\mp iM\sqrt{\frac{2m\omega}{1-q^{-2M}}}\,x}
\,\left(p\pm \sqrt{\frac{2m\omega}{1-q^{-2M}}}\right) \right.\nn\\
&&\hspace{40mm}\left. + \,\left(p \pm \sqrt{\frac{2m\omega}{1-q^{-2M}}}\right)
\,q^{\frac{i}{2}M(xp+px)\pm iM\sqrt{\frac{2m\omega}{1-q^{-2M}}}\,x}\right]\nn
\eea
This Hamiltonian has eigenfunctions with the eigenvalues (\ref{e6})
and (\ref{e9}). For the spectrum (\ref{e6}) we know the eigenfunctions 
explicitly. We discuss the ground states (\ref{e16}):
\bea
|0\rangle^{(M)}_{(\sigma,s,\mu)}&=& \;  \sum_{n=-\infty}^{+\infty} 
  \tilde{c}^{(\sigma,s,\mu)}_n\,|nM+\mu,\sigma\rangle |s\rangle \label{f12}\\
&=&\; \sum_{n=-\infty}^{+\infty}\left(\pm\frac{\sigma}{s}\, 
\sqrt{\frac{2m\omega}{1-q^{-2M}}}\right)^n\,q^{-\frac{1}{2}(Mn^2+Mn+2\mu n)}
\,c_0^{(\mu)} |nM+\mu,\sigma\rangle |s\rangle  \nn
\eea
As we are dealing with reducible representations of the algebra (\ref{e1})
where we express $P$, $X$ and $U$ in $p$ and $x$ as in (\ref{f1}) and 
(\ref{f3}) we have to label the irreducible parts by the variable $s$. 
We follow the notation of ref. \cite{pha}.
The sign factor $\pm\sigma$ again refers to the two choices in (\ref{f8}).\\
The eigenstates of $P$ now have to be expressed in a basis which is labeled
by the eigenvalues of $p$ in (\ref{f3}):
\be
|n,\sigma\rangle |s\rangle = 
 \int dp_0 \, q^{\frac{n}{2}}\, \delta\left(p_0 \pm 
\sqrt{\frac{2m\omega}{1-q^{-2M}}} -\sigma s q^n\right)\: |p_0\rangle 
\label{f13}
\ee
and we obtain
\bea
|0\rangle^{(M)}_{(\sigma,s,\mu)}
&=&\sum_{n=-\infty}^{+\infty}{\textstyle \left(\pm\frac{\sigma}{s} 
\sqrt{\frac{2m\omega}{1-q^{-2M}}}\right)}^n\,
q^{-\frac{1}{2}(Mn^2+2\mu n-\mu)}\,c_0^{(s,\mu)} \label{f14}\\
&&\hspace{15mm} \cdot \int dp_0 \, \delta\left(p_0 \pm 
\sqrt{\frac{2m\omega}{1-q^{-2M}}}-\sigma s q^{nM+\mu}\right) \;|p_0\rangle \nn
\eea
These eigenfunctions live on a q-deformed lattice in momentum space. 
For \linebreak
$\sigma=+1$, $p_0$ ranges from $\mp \sqrt{\frac{2m\omega}{1-q^{-2M}}}$ to
$+\infty$, for $\sigma=-1$, $p_0$ ranges from $-\infty$ to 
$\mp \sqrt{\frac{2m\omega}{1-q^{-2M}}}$.\\[2mm]
The corresponding wave functions for the ground states in $x$-space are:
\bea
{\Psi_0}^{(M)}_{(\sigma,s,\mu)}\,(x)&=&\sum_{n=-\infty}^{+\infty}{\textstyle 
\left(\pm\frac{\sigma}{s}\sqrt{\frac{2m\omega}{1-q^{-2M}}}\right)}^n\,
q^{-\frac{1}{2}(Mn^2+2\mu n-\mu)}\,c_0^{(s,\mu)} \label{f15}\\
&&\hspace{22mm} \cdot\,e^{ix\,( \sigma s q^{nM+\mu}\mp
\sqrt{\frac{2m\omega}{1-q^{-2M}}}\,)} \nn
\eea
They can be seen to be eigenfunctions of the Hamiltonian (\ref{f11}) without 
ever referring to q-deformation. The excited states can be obtained by 
applying the $a^+$-operators expressed in terms of $x$ and 
$p\!=\!-i\frac{\partial}{\partial x}$ to the ground state wave functions. \\
It is interesting to see the limit $q\rightarrow 1$ on the wave functions,
this will be done in the next chapter.\\

\section{The wave functions in the limit $q\rightarrow 1$}

\setcounter{equation}{0}

The q-deformation of the Hamiltonian has led to a degeneracy of the 
spectrum that is not encountered for $h=0$. It is interesting to see
how the wave functions of the q-deformed oscillator behave for 
$h\rightarrow 0$.
First we realize that for $h\rightarrow 0$ $E_{[\infty]} \rightarrow \infty$.
The part $E> E_{[\infty]}$ disappears from the spectrum.\\
To study the behaviour of the wave functions for $E< E_{[\infty]}$, we
start with the ground state (\ref{e12}). To simplify the discussion, we
consider the case $M=1$ first.
\be
|0\rangle^{(1)}_{(\sigma)} =
 \sum_{n=-\infty}^{+\infty}\left(-\sigma\frac{\alpha}{\beta}\,\right)^n\,
q^{-\frac{1}{2}(n^2+n)}
\,c_0\; |n,\sigma\rangle 
\label{g1}
\ee
The normalization constant $c_0$ can be computed:
\bea
|c_0|^{-2} &=& \sum_{n=-\infty}^{+\infty}\left(\frac{\alpha}{\beta}\,
 \right)^{2n} \,q^{-n^2-n} \label{g2} \\
&=&  \sum_{n=-\infty}^{+\infty} e^{\textstyle -h(n^2+n) 
  + n \,{\rm ln}\frac{2mw}{1-q^{-2}}} \nn
\eea
This expression will be singular for $h\rightarrow 0$ and we want
to estimate its behaviour. We complete the square in the exponent of 
(\ref{g2}) and obtain:
\be
|c_0|^{-2} = e^{\frac{h}{4}}\, e^{\textstyle -\frac{1}{2} 
 \,{\rm ln}\frac{2m\omega}{1-q^{-2}} 
 + \frac{1}{4h}\left({\rm ln}\frac{2m\omega}{1-q^{-2}}
  \right)^2}\,\sum_{n=-\infty}^{+\infty}
   e^{\textstyle -h\left(n+\frac{1}{2}
 -\frac{1}{2h}\,{\rm ln}\frac{2m\omega}{1-q^{-2}} \right)^2}\nn
\label{g3}
\ee
The sum in (\ref{g3}) can be estimated by considering it to approach the
upper and lower Riemann integral of a Gaussian. We obtain
\be
\sum_{n=-\infty}^{+\infty} e^{-h(n+a)^2}\; \longrightarrow \;
  \sqrt{\frac{\pi}{h}}
\label{g4}
\ee
independent of $a$. This now exhibits the behaviour of $c_0$ for 
$h\rightarrow 0$.\\
For any state
\be
|f\rangle = \sum_{n=-\infty}^{+\infty}\,c_n\,|n,+\rangle \;\;\; ,
\label{g5}
\ee
the norm will approach an integral in the limit $h\rightarrow 0$.
\be
\langle f|f\rangle = \sum_{n=-\infty}^{+\infty} |c_n|^2
\label{g6}
\ee
We identify the points of the "lattice", according to (\ref{f3}), with points
$p$ in $R^1$:
\be
\sigma q^n = p + \frac{1}{\sqrt{1-q^{-2}}}\,\gamma \;
        =\; p + \sqrt{\frac{2m\omega}{1-q^{-2}}}
\label{g7}
\ee
For the sake of definiteness we have chosen the positive sign for $\gamma$
in (\ref{f8}). We see that the variable $p$ will range from 
$-\sqrt{\frac{2m\omega}{1-q^{-2}}}$ to $+\infty$, if $\sigma=+1$. If  
$\sigma=-1$, $p$ will range from $-\infty$ to 
$-\sqrt{\frac{2m\omega}{1-q^{-2}}}$. In the limit
$h\rightarrow 0$, the support of the wave function for $\sigma=-1$ will
disappear, for $\sigma=+1$ it will range from $-\infty$ to $+\infty$. \\
If we had chosen the other sign for $\gamma$ in (\ref{f8}) 
the roles of the $\sigma=+1,-1$ states would have been interchanged. 
This shows how the $\sigma$-degeneracy of the states disappears.

Let us now transform (\ref{g6}) into an integral with the identifications
(\ref{g7}) for \linebreak 
$\sigma=+1$. We have:
\bea
q^n &=& p + \sqrt{\frac{2m\omega}{1-q^{-2}}}\hspace{4mm},\hspace{10mm}
n\, =\, \frac{1}{h} \,{\rm ln}\left(p+\sqrt{\frac{2m\omega}{1-q^{-2}}}\right) 
\label{g8}\\[1mm]
\Delta_n p &=& q^{n+1} - q^n \,=\, (q-1)\,q^n \,=\, 
 (q-1)\,\left(p+\sqrt{\frac{2m\omega}{1-q^{-2}}}\right) \nn
\eea

We find:
\be
\langle f|f\rangle = \sum_{n=-\infty}^{+\infty}\, |c_n|^2 \,
 \frac{\Delta_n p}{(q-1)\left(p+\sqrt{\frac{2m\omega}{1-q^{-2}}}\right)}
\;\longrightarrow \;\int_{-\infty}^{+\infty}\,|f(p)|^2 \,dp
\label{g9}
\ee
where\\[-3mm]
\be
f(p) = \frac{c_n}{\sqrt{(q-1)\left(p+\sqrt{\frac{2m\omega}{1-q^{-2}}}\right)}}
\hspace{4mm},\hspace{8mm} 
n = \frac{1}{h} \,{\rm ln}\left(p+\sqrt{\frac{2m\omega}{1-q^{-2}}}\right) 
\label{g10}
\ee
This is the formula by which we identify the wave function of the ground state 
in the limit $h\rightarrow 0$.\\ 
We now go back to (\ref{g1}) and make the substitutions of (\ref{g10}) on
the indiviual parts:
\bea
q^{ -\frac{1}{2}n} &=& e^{\textstyle -\frac{1}{2}\left({\rm ln}
 \sqrt{\frac{2m\omega}{1-q^{-2}}} +
   {\rm ln}\left(1+p\sqrt{\frac{1-q^{-2}}{2m\omega}}\right)\right)} \nn\\[2mm]
q^{ -\frac{1}{2}n^2} &=& e^{\textstyle -\frac{1}{2h}\left({\rm ln}
 \sqrt{\frac{2m\omega}{1-q^{-2}}} +
   {\rm ln}\left(1+p\sqrt{\frac{1-q^{-2}}{2m\omega}}\right)\right)^2}
\label{g11}\\[2mm]
{\left(-\frac{\alpha}{\beta}\right)}^n&=&e^{\textstyle \frac{1}{h}\,{\rm ln}
 \sqrt{\frac{2m\omega}{1-q^{-2}}}\left({\rm ln}\sqrt{\frac{2m\omega}{1-q^{-2}}}
 + {\rm ln}\left(1+p\sqrt{\frac{1-q^{-2}}{2m\omega}}\right)\right)} \nn\\
c_0 &=& {\left(\frac{h}{\pi}\right)}^{\frac{1}{4}}\,
  e^{\textstyle -\frac{h}{8}+ \frac{1}{2}\,{\rm ln}
 \sqrt{\frac{2m\omega}{1-q^{-2}}} 
  - \frac{1}{2h} \left({\rm ln}\sqrt{\frac{2m\omega}{1-q^{-2}}}\right)^2} \nn
\eea
Putting all together we find the surprising result:
\be
f(p)= \frac{1}{\textstyle (m\omega\pi)^{\frac{1}{4}}} \,
   e^{\textstyle -\frac{p^2}{2m\omega}}
\label{g12}
\ee
This is the correctly normalized ground state wave function of the 
undeformed harmonic oscillator.\\
It is easy to see that for $M\ne 1$ all the different wave functions for the
ground state collapse to the same wave function (\ref{g12}). Thus all the 
degeneracy is removed in the limit $h\rightarrow 0$.\\
The excited states are obtained from the ground states by applying creation
operators. As these operators tend to the undeformed creation operators of
the harmonic oscillator we reproduce all the wave functions of the
undeformed oscillator. It is not surprising that a direct calculation on the
states defined by the coefficients (\ref{e25}) leads to the same result.

\section{q-deformed Hermite polynomials}

\setcounter{equation}{0}

We recall the definition of the ground state ($M=1$)
\be
a\,|0\rangle = \left( \alpha\, U^{-2} + \beta \,U^{-1}P\right) |0\rangle 
\,=\, 0
\label{h1}
\ee
with the immediate consequence:
\be
U\, |0\rangle = -\,\frac{\alpha}{q\,\beta}\,P^{-1}\, |0\rangle
\label{h2}
\ee
This relation together with the definition of the algebra (\ref{e1}) can be
used to show:
\be
a^+\, |0\rangle \,=\,  \frac{i}{q^{\frac{1}{2}}\beta}\,X\, |0\rangle 
\label{h3}
\ee
The ground state as well as the one-particle state can be obtained by 
applying a polynomial in $X$ to the groundstate. We are going to show
that this is true for the n-particle state as well:
\be
\left(a^+\right)^n \,|0\rangle \;=\; \left(\frac{1}{\sqrt{2}}\right)^n
H_n^{(q)}\!\left(\frac{i X}{\sqrt{2}\,\beta}\right)\,|0\rangle
\label{h4}
\ee
$H_n^{(q)}$ will turn out to be a polynomial of degree $n$, satisfying 
a q-deformed recursion formula for Hermite polynomials. The normalization
and the scaling of the argument to a dimensionless parameter
\be
\xi \,=\, \frac{i X}{\sqrt{2}\,\beta} \,=\, \sqrt{m\omega}\,X
\label{h5}
\ee
are chosen to identify $H_n^{(1)}(\xi)$ with the usual Hermite polynomial.

To prove eqn. (\ref{h4}) we start from the following relation which can
be obtained from the definition of the algebra (\ref{e1})
\be
a^+ \xi \,=\, q^{-2}\, \xi a^+ -\,\frac{1}{\sqrt{2}} \,q^{-\frac{3}{2}}
\label{h6}
\ee
and we prove by induction:
\be
\left( \left(a^+\right)^{n+1} - q^{-\frac{1}{2}}q^{-2n}\,\sqrt{2} \,
\xi \,\left(a^+\right)^n + q^{-2}\,
\frac{1-q^{-2n}}{1-q^{-2}} \,\left(a^+\right)^{n-1} \right) |0\rangle \, =\, 0
\label{h7}
\ee
For $n=0$, this is just equation (\ref{h3}). With the help of (\ref{h6}) the
step $n$ to $n+1$ is easily verified.\\
Eqn. (\ref{h7}) shows that $H_n^{(q)}(\xi)$ is a polynomial of degree $n$ in 
$\xi$. Eqn. (\ref{h7}) can now be written 
in the form
\be
\left(H_{n+1}^{(q)}(\xi) - q^{-\frac{1}{2}}q^{-2n} \,2\, \xi\:H_n^{(q)}(\xi) 
+ 2 q^{-2}\,\frac{1-q^{-2n}}{1-q^{-2}}\: H_{n-1}^{(q)}(\xi)\right)\,|0\rangle
\,=\, 0
\label{h8}
\ee
If a polynomial satisfies a recursion formula
\be
H_{n+1}^{(q)}(\xi) - q^{-\frac{1}{2}}q^{-2n} \,2\,\xi\:H_n^{(q)}(\xi) 
+ 2 q^{-2}\,\frac{1-q^{-2n}}{1-q^{-2}}\: H_{n-1}^{(q)}(\xi)\,=\, 0
\label{h9}
\ee
and if
\be
H_0^{(q)}(\xi) \,=\, 1 \hspace{5mm},\hspace{9mm}
H_1^{(q)}(\xi)\,=\, 2\,q^{-\frac{1}{2}}\xi
\label{h10}
\ee
then $H_{n+1}^{(q)}(\xi)$, obtained from (\ref{h9}), will be the correct 
polynomial in (\ref{h4}).\\
For $q=1$, (\ref{h9}) is just the recursion formula for Hermite polynomials
and (\ref{h10}) are the first two Hermite polynomials.

We found that the states of the spectrum of the q-deformed harmonic oscillator
for $E<E_{[\infty]} = \omega(1-q^{-2})^{-1}$  can be obtained by applying 
q-deformed Hermite polynomials to the ground states $|0\rangle_{(\sigma=+1)}$
and $|0\rangle_{(\sigma=-1)}$. However, as indicated in (\ref{e11}), these 
states do not form a complete set of states in the 
Hilbert space where $P$ and $X$ are represented. This is in contrast to the 
case $q=1$.\\
The orthogonality relations of the eigenstates of the Hamiltonian lead to
interesting orthogonality relations for the q-deformed Hermite polynomials.
They will involve the q-deformed cosine and sine functions as the ground state
in (\ref{h4}) will have to be expressed in the $X$-basis.\\

\subsection*{Acknowledgement}
\vspace{-5mm}
We would like to thank J. Seifert and M. Aita for discussions and helpful
remarks.

\end{document}